\documentclass[final,5p,times,twocolumn]{elsarticle}

\usepackage{amssymb}

\usepackage{amsmath}

\journal{ }

\begin{document}

\begin{frontmatter}



\title{On two-dimensional exciton bound by distant ionized-donor in a narrow quantum well}


\author{M.~Tytus}
\author{W.~Donderowicz}
\author{L.~Jacak}

\address{Institute~of~Physics, Wroc\l{}aw~University~of~Technology, Wybrze\.ze~Wyspia\'nskiego 27, 50-370~Wroc\l{}aw, Poland}

\begin{abstract}
The ground state energy of exciton bound by distant ionized donor impurity
in two-dimensional semiconductor quantum well (QW)
is studied theoretically within the Hartree approach in the effective mass approximation.
The influence 
of the distance between QW plane and ionized donor,
as well as
of the electron-hole mass ratio,
the magnetic field 
and
dielectric constant of the barrier material
on the stability of exciton bound by ionized donor impurity
is analyzed and discussed.
\end{abstract}




\end{frontmatter}


\section{Introduction}

One of the simplest possible three-particle bound-exciton complex $(D^+,X)$ 
consist of an exciton $X$ (electron hole pair) 
bound to an ionized donor impurity $D^+$.
Its possible existence was first predicted by Lampert~\cite{Lampert} in 1958.
Since then its stability and binding energy
as a function of the electron to hole effective mass ratio $\sigma=m_e/m_h$
has been the subject of few theoretical studies 
in bulk (3D) semiconductor \cite{Skettrup, Stauffer} and 
in two-dimensional QW structures \cite{Stauffer,Ruan}.
As a result in 3D case Stauffer and St\'eb\'e using a 55-term Hylleraas-type wave function 
obtained \cite{Stauffer} that the complex $(D^+,X)$ is stable until 
$\sigma<\sigma_c^{3D}=0.365$.
Skettrup \textit{et al.} using more elaborated wave functions 
obtained \cite{Skettrup} critical value of this ratio $\sigma_c^{3D}=0.426$. 
In 2D case the overlappings between the wave functions 
of the constituents of the $(D^+,X)$ complex 
become more important  (due to the quantum confinement), 
binding energy of the complex is increased
and so the stability in 2D structures is increased 
compared to their 3D counterparts.
Therefore it is expected that the observation of bound excitons 
should be easier in 2D structures than in the bulk. 
In the case of two-dimensional QW, 
Stauffer and St\'eb\'e using the same method as in 3D case
obtained \cite{Stauffer} critical value $\sigma_c^{2D}=0.88$.
However more recently, 
Ruan and Chang using the hyperspherical adiabatic expansion approach
found that the complex $(D^+,X)$ is stable in 2D case with any value 
of the electron to hole effective mass ratio in the range $0\leq\sigma\leq 1$
whereas for $1\leq\sigma\leq \infty$ negatively charged acceptor ion $A^-$ 
can bound exciton and the complex $(A^-,X)$ is stable \cite{Ruan}.
%
%

For finite well width 
Liu and co-workers \cite{Liu3}, 
using a two-parameter wave function,
calculated variationally 
the binding energy of an exciton 
bound to an ionized donor impurity $(D^+,X)$ 
in GaAs/Al$_x$Ga$_{1-x}$As QW
for the values of the well width from 1 to 30$\,$nm,
when the dopant is located in the center of the well 
and at the edge of the well. 
%
%
da Cunha Lima \textit{et al.} \cite{Lima} 
performed a variational calculation of the binding energy of the $(D^+,X)$ complex 
for all values of well widths, 
and impurity position inside the well,
including $\Gamma$-$X$ mixing in GaAs/AlAs QWs.
St\'eb\'e and co-workers \cite{Stebe} studied variationally  
the influence of the magnetic field on the binding energy of $(D^+,X)$ 
in GaAs/Al$_x$Ga$_{1-x}$As QWs.
%
%

Nevertheless no one has yet analyzed the impact of 
ionized donor shifted from the QW plane.
The lateral crossection of singular potential of ionized donor, 
as acting on charge carriers in distant 2D well, 
resembles a nonsingular potential of type-II quantum dot (defined by the electrostatic field) 
--- thus recognition of exciton evolution with respect to the donor separation 
is of crucial importance in order to differentiate both confinements. 

In this paper the ground state energy of exciton 
bound by ionized donor shifted from two-dimensional QW
is studied theoretically within the Hartree approach 
in the effective mass approximation.
The influence of 
the donor distance,
the electron-hole mass ratio,
varying dielectric constant of barrier material
and of the external uniform magnetic field (aligned across the QW plane)
on the stability 
of two-dimesional exciton bound by ionized donor
is analyzed and discussed.

\section{Model and method}

For the model analysis, we assume 
that the QW is quasi-two-dimensional 
and lies in the x-y plane,
while
the magnetic field
is aligned across this plane, i.e., along the z axis.
Moreover,
we restrict our model 
only to the spatial coordinates
--- spin degrees of freedom 
and the associated Zeeman splitting (linear in B)
were not included in our description
(for GaAs this splitting is very small $\sim 0.03\,$meV/T.)

In the QW plane 
potential of ionized donor shifted by the distance $d$ 
in the axial direction has the form
\begin{equation}
  \label{Vd}
    V_i\left(\rho_i\right) = \mp \frac{q^2}{4\pi\epsilon_1\epsilon_0}\frac{1}{\sqrt{\rho_i^2+d^2}},
\end{equation}
where minus sign corresponds to the electron ($i=e$), plus to the hole ($i=h$),
$\rho_e$ and $\rho_h$ are the radial distances of electron and hole in the QW plane,
$q$ is the elementary positive charge
and $\epsilon_1$ is the relative dielectric constant of the barrier material.

Despite the fact that the potential~\eqref{Vd}
is attractive only for one type of charge carrier
it is however possible for the distant donor to captured the electron-hole pair (exciton) 
due to Coulomb interaction between charge carriers.

It should be noted here, 
that in more realistic model, for very small $d$ and shallow QWs,
electron tunneling through the potential barrier has to be taken into account.
Nevertheless, the probability of this process rapidly decreases 
as the donor is shifted away from QW plane.

Within the Hartree method exact exciton wave function can be approximated by
$\Psi\left(\textbf{r}_e,\textbf{r}_h\right) = \psi_e\left(\textbf{r}_e\right) \psi_h\left(\textbf{r}_h\right)$
where $\textbf{r}_i=(\rho_i,\varphi_i)$ and $i=e,h$.
Using axial symmetry we assume one particle wave functions in the form 
\begin{equation*}
     \psi_s\left(\textbf{r}_s\right)=\frac{1}{\sqrt{2\pi}}\exp\left(il_s\varphi_s\right)\phi_s\left(\rho_s\right),
\end{equation*}
where $l_s=0,\pm 1,\pm 2,\ldots$ and $s=e,h$.
Then the single-particle Hartree energies and wave functions are found 
in the effective mass approximation
by iterative solving of self-consistent Hartree equations
\begin{align}
\label{RS}
\begin{split}
		\Bigg[
	  -\frac{\hbar^2}{2m_e}\frac{1}{\rho_e}\frac{\partial}{\partial \rho_e}\left(\rho_e \frac{\partial}{\partial \rho_e}\right) +\frac{\hbar^2}{2m_e}\frac{l_e^2}{\rho_e^2} +&\\ 
		U_e\left(\rho_e\right) +\frac{l_e}{2}\hbar\omega_{ce} 
		&\Bigg] 
		\phi_e\left(\rho_e\right) = \varepsilon_e \phi_e\left(\rho_e\right),\\
		\Bigg[
		-\frac{\hbar^2}{2m_h}\frac{1}{\rho_h}\frac{\partial}{\partial \rho_h}\left(\rho_h \frac{\partial}{\partial \rho_h}\right) +\frac{\hbar^2}{2m_h}\frac{l_h^2}{\rho_h^2} +&\\ 
		U_h\left(\rho_h\right) -\frac{l_h}{2}\hbar\omega_{ch}
		&\Bigg] 
		\phi_h\left(\rho_h\right) = \varepsilon_h \phi_h\left(\rho_h\right),
\end{split}
\end{align}
with the effective Hartree patentials
\begin{equation}
\label{Ue}   
	 U_e\left(\rho_e\right)  =  V_e\left(\rho_e\right)  +\frac{1}{8}m_e\omega_{ce}^2\rho_e^2 
    -\frac{q^2}{4\pi\epsilon_2\epsilon_0} \int \frac{\left|\psi_h\left(\textbf{r}_h\right)\right|^2}{\left|\textbf{r}_e-\textbf{r}_h\right|}d\textbf{r}_h,
\end{equation}  
\begin{equation}
	\label{Uh}   
   U_h\left(\rho_h\right)  =  V_h\left(\rho_h\right)  +\frac{1}{8}m_h\omega_{ch}^2\rho_h^2 
    -\frac{q^2}{4\pi\epsilon_2\epsilon_0} \int \frac{\left|\psi_e\left(\textbf{r}_e\right)\right|^2}{\left|\textbf{r}_e-\textbf{r}_h\right|}d\textbf{r}_e,
\end{equation}
where $m_e$ and $m_h$ are effective electron and hole masses respectively,
$\omega_{ce}=qB/m_e$ and $\omega_{ch}=qB/m_h$ are electron and hole cyclotron frequencies
and $\epsilon_2$ is the relative dielectric constant of the QW material.

The exciton energy in Hartree approximation is given by
\begin{equation}
\label{Ex}
		E = \varepsilon_e + \varepsilon_h - V_C,
\end{equation}
where
\begin{equation}
\label{Vc}
		V_C  =  -\frac{q^2}{4\pi\epsilon_2} \iint \frac{\left|\psi_e\left(\textbf{r}_e\right)\right|^2\left|\psi_h\left(\textbf{r}_h\right)\right|^2}{\left|\textbf{r}_e-\textbf{r}_h\right|}d\textbf{r}_e d\textbf{r}_h.
\end{equation}
%
%
As we deal with single electron-hole pair 
there is no exchange energy term 
(related to Pauli exclusion principle)
and only correlation energy is omitted.
Moreover,
as it was shown for quantum dots 
(whose potential is similar to the potential of shifted donor in QW plane),
the contribution of the correlation 
to the total energy for single electron-hole pair 
is expected to be less than 2\% \cite{Brasken}.

Hartree equations \eqref{RS} 
were solved numerically with finite difference scheme on nonuniform grid
(more details about the implementation of this finite difference scheme 
can be found in work of Peeters \textit{et al.} \cite{Peeters}).
Using this scheme we obtained symmetric tridiagonal matrix.
Its eigen values were calculated with Martin-Dean algorithm \cite{Dean},
whereas eigen vectors were found using DWSZ method \cite{Dy}.
Hartree integrals in \eqref{Ue}, \eqref{Uh} and \eqref{Vc}
were calculated with use of logarithmically weighted method 
after Janssens \textit{et al.} \cite{Janssens}.
The convergence in the self-consistent Hartree procedure 
is obtained in a few rounds.

\section{Results and discussion}

\begin{figure}[tb]
	\centering
	\includegraphics{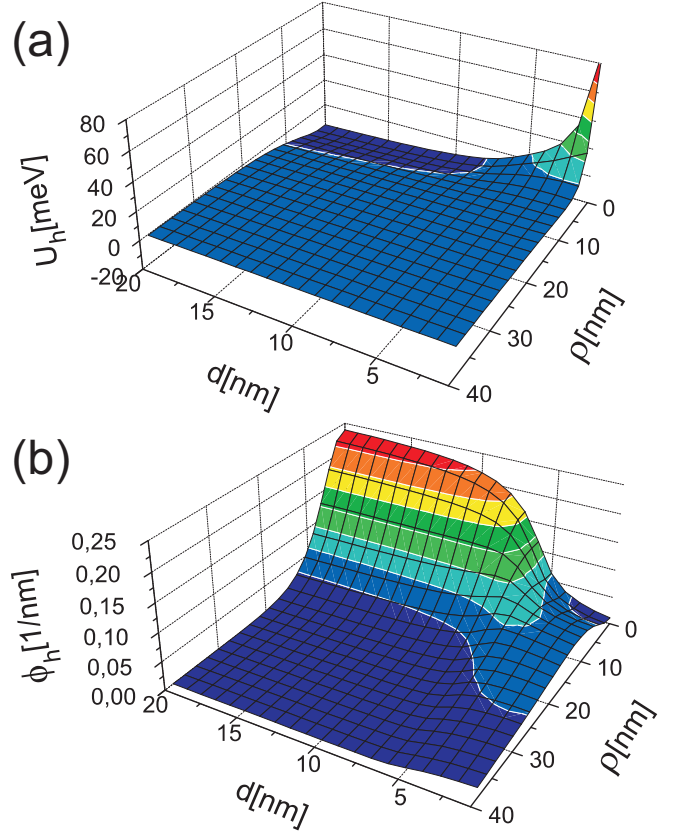} 
	\caption{\label{fig:Uh_Yh_d}The effective Hartree hole potential
	                            as a function of the donor distance $d$ from the QW plane (a)
	                            and corresponding hole wave function (b)
	                            for $\sigma=0.176$ and $\epsilon_1=\epsilon_2=12.4$.
	        }
\end{figure}

As an example
we consider the case of GaAs semiconductor QW
and choose 
material parameters
$\epsilon_2=12.4$, $m_e=0.0665$, $m_h=0.3774$ so that $\sigma=0.176$.
For such choice 
Fig. \ref{fig:Uh_Yh_d} shows dependence of
effective Hartree hole potential~(a) and 
corresponding hole wave function~(b)
on the donor distance $d$ from the plane
for listed parameters.
For the barrier material, at this point, 
we choose $\epsilon_1=\epsilon_2$.
When~$d$ is small 
Hartree hole potential is repulsive in the center of the QW
and has minimum at certain distance.
As~$d$ increases donor repulsion become weaker
and for $d\sim 5\,$nm hole moves to the center.

Electron Hartree potential
does not change qualitatively 
as the donor is moved away from the plane 
--- first becomes shallower with increasing $d$,
next it deepens when the hole moves to the center 
(due to the increase of electron-hole coulomb interaction),
and then becomes shallower again.

\begin{figure}[tb]
	\centering
	\includegraphics{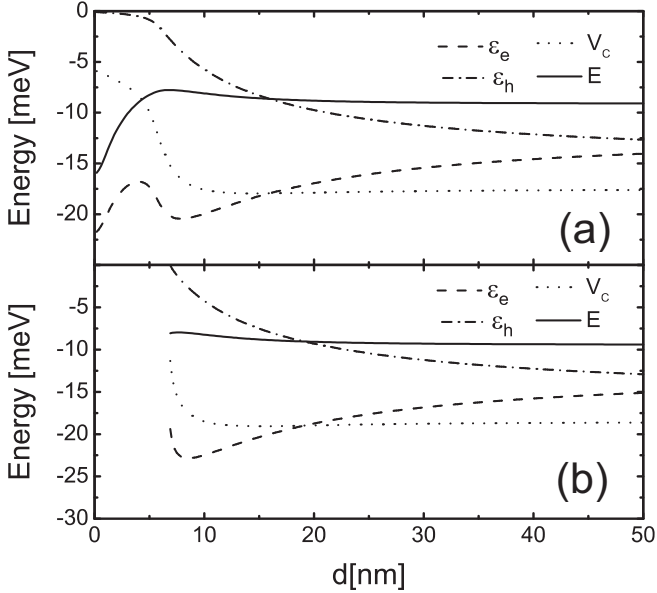} 
	\caption{\label{fig:E_d}Electron (dashed line) and hole (dash-dot) Hartree energies,
													Coulomb (dot) and exciton (solid) energies
	                        as a function of the donor distance $d$ from the QW plane for
	                        $\epsilon_1=\epsilon_2=12.4$ (a)
	                        and $\epsilon_1=10.1$, $\epsilon_2=12.4$ (b).
	        }
\end{figure}

Fig. \ref{fig:E_d}(a) illustrates
electron (dashed line) and hole (dash-dot) Hartree energies
as well as Coulomb (dot) and exciton (solid) energies
as a function of the donor distance $d$ from the QW plane
also for 
$\epsilon_1=\epsilon_2$.
%
%
When $d$ is small
repulsive potential of ionized donor in the center of the well 
is bigger than the electron attraction
and
the hole is located in the ring around the axis of symmetry (cf. Fig. \ref{fig:Uh_Yh_d}(b)).
Therefore, Coulomb attraction grows slowly with $d$, 
and the exciton energy increases
mainly due to the increase of electron energy in ionized donor potential.
%
%
For bigger $d$
donor repulsion becomes weaker 
and the hole moves to the center.
Electron and hole Hartree energies together
decreases slightly faster than the Coulomb energy
so the energy of exciton also decreases.

For distance greater than about 15$\,$nm
exciton energy almost stop changing
despite further change of electron and hole oneparticle Hartree energies.
(It should be noted here 
that in the system of interacting particles 
unambiguous qualification of oneparticle energies is not possible.
The energy of the whole system
is the only direct contact with the experiment, 
therefore, further change of oneparticle energies 
may be irrelevant from a physical point of view.)
So for large distances~$d$ 
energy should correspond to the energy of free two-dimensional exciton.

On the other hand,
as can be shown
by extrapolating for small $d$ the Hartree hole energy
(it never reaches zero)
ionized donor should bound exciton even for $d=0$.

The other situation is for $\epsilon_1<\epsilon_2$.
As an example we will consider GaAs/AlAs for which
$\epsilon_1=10.1$ and $\epsilon_2=12.4$.
In contrast to the previous situation, 
as we can see in Fig. \ref{fig:E_d}(b), 
the hole is not bound
until the distance of donor
reaches the critical value $d_{min}$.
Thus the lower dielectric constant of the barrier,
prevents the exciton binding for small $d$
--- hole is more strongly repelled
and the electron Coulomb attraction is insufficient to bound a hole
until we move the donor at an appropriate distance.
The question arises
how far to move the donor from the plane of the well
for a given dielectric constant of the barrier
in order to obtain a bound state.

\begin{figure}[tb]
	\centering
	\includegraphics{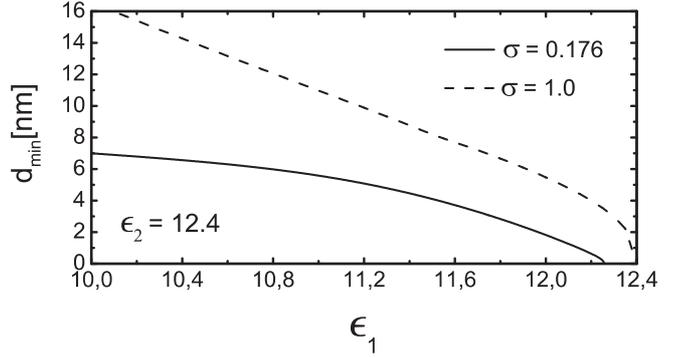} 
	\caption{\label{fig:dmin}The dependence of critical distance $d_{min}$ on the $\epsilon_1$ 
													 for $\sigma=0.176$ (solid line) and for $\sigma=1.0$ (dashed line).
					}
\end{figure}

Fig. \ref{fig:dmin} shows 
the dependence of this critical distance
on the $\epsilon_1$ (solid line) for GaAs/Al$_x$Ga$_{1-x}$As,
for which we can assume that 
\begin{equation}
\label{e1_x}
	\epsilon_1(x)=12.4-2.3x.
\end{equation}	
As one might expect the critical distance $d_{min}$ 
decreases with increasing $\epsilon_1$
and reaches zero before $\epsilon_1$ equals $\epsilon_2$.
It follows that the exciton bound state
could exists for $\epsilon_1=\epsilon_2$.
%

The situation is different for mass ratio $\sigma=1$ (dashed line in Fig. \ref{fig:dmin}), 
for which $d_{min}$ tends to zero when $\epsilon_1$ tends to $\epsilon_2$.
Although we can not show that $d_{min}$ 
is exactly equal to zero for $\epsilon_1=\epsilon_2$,
however, by our calculation
we can show that it is smaller than the desired numerical accuracy.


So far our approach was limited to
the analysis of the Hartree hole energy
--- if it was less than zero
we assumed that exciton is bound by ionized donor impurity.
Nevertheless, even in this situation,
created complex may be unstable 
due to the following dissociation processes
\begin{align}
\label{disD}
	(D^+,X) \rightarrow & D^0 + h,\\
\label{disX}
	(D^+,X) \rightarrow & D^+ + X.
\end{align}
In these equations
$(D^+,X)$ and $D^0$ denote respectively exciton or electron bound by ionized donor in the QW,
while
$h$ and $X$ denote free hole and free exciton in the QW plane.
Therefore, 
we need to consider the 
binding energies
%
%
$E^B_{D^0} = E_{D^0} + E^f_h - E_{(D^+,X)}$,
$E^B_X     = E^f_X - E_{(D^+,X)}$
%
%
whose physical meaning is that
the $E^B_{D^0}$ is the minimum energy 
required to liberate the hole from the bound exciton
and $E^B_X$ is the minimum energy required 
to liberate the exciton from the influence of ionized donor.
So the complex remains stable if $E^B_{D^0}>0$ and $E^B_X>0$.

In our calculations
for $E^f_X$ we take the value to which the energy of bound exciton tends, when $d$ tends to infinity,
and we put $E^f_X$ equal $\hbar\omega_{ch}/2$ (lowest landau level in magnetic field).

\begin{figure}[tb]
	\centering
	\includegraphics{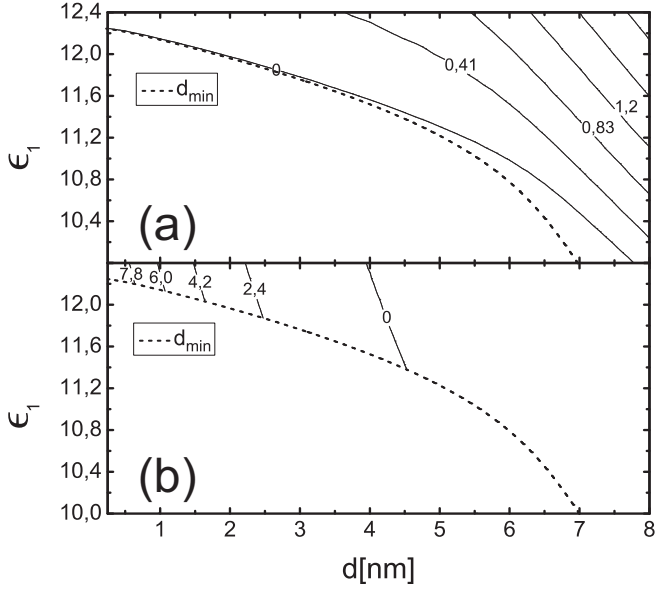} 
	\caption{\label{fig:Eb_ed_d}Contour plots of energies $E^B_{D^0}$ (a) and $E^B_X$ (b), 
														  both in meV,			
					                    depending 
												      on the dielectric constant of barrier material $\epsilon_1$ and 
												      the donor distance $d$ from the QW plane ($\sigma=0.176$).
													    Dotted lines indicate the limit distance for donor.
	        }
\end{figure}

Fig. \ref{fig:Eb_ed_d} represents
the dependence of the energies $E^B_{D^0}$ and $E^B_X$, 
both in meV,
on the donor distance from the QW and
on the dielectric constant of the barrier material
in the range $10<\epsilon_1<12.4$ (as for Al$_x$Ga$_{1-x}$As).
Additional dotted lines indicate 
the limit distance for donor (cf. Fig. \ref{fig:dmin}).
As can be seen by comparing parts (a) and (b) 
the complex is stable when $11.4\leq\epsilon_1\leq 12.4$
(which incidentally corresponds to the interval in which our heterostructure has a direct energy gap)
and only for $d_{min}<d\lesssim 4\,$nm
(for $\epsilon_1$ close to $\epsilon_2$ we should be especially careful 
with the interpretation of our results  
because the difference between energy gaps of GaAs/Al$_x$Ga$_{1-x}$As becomes small
so the electron can tunnel through the potential barrier).
For $d$ greater than $\sim 4\,$nm,
in the whole range of $\epsilon_1$,
complex may dissociate into $X$ and $D^+$
or,
if in addition $d$ is not much larger than the $d_{min}$,
into hole and $D^0$.

\begin{figure}[tb]
	\centering
	\includegraphics{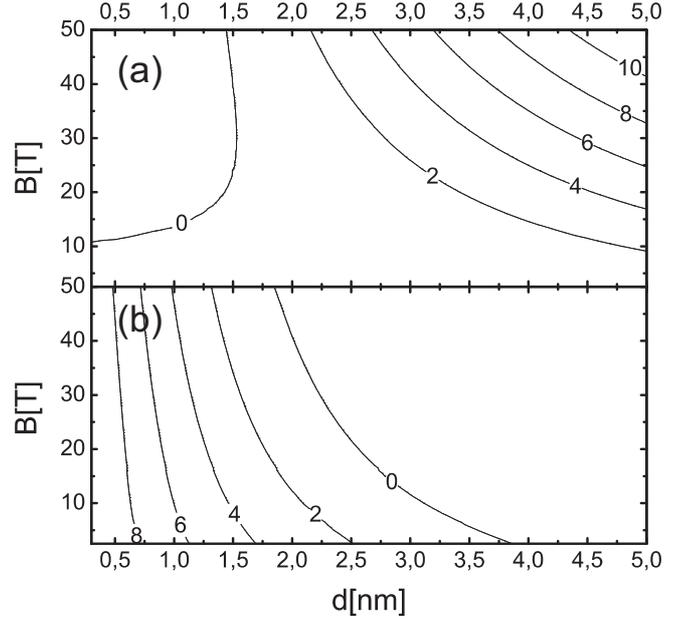} 
	\caption{\label{fig:Eb_B_d}The dependence of energies $E^B_{D^0}$ (a) and $E^B_X$ (b),
	                           both in meV, 
														 on the donor distance $d$ from the plane of QW
														 and magnetic field $B$ 
													   for $\sigma=0.176$ and $\epsilon_1=11.71$, $\epsilon_2=12.4$.
	        }
\end{figure}

It is also interesting to check 
what impact on the stability of discussed complex 
has the magnetic field 
perpendicular to
the QW plane.
Fig. \ref{fig:Eb_B_d} shows 
the dependence of the binding energies $E^B_{D^0}$ and $E^B_X$
on the donor distance from the QW 
and  magnetic field
for $\sigma=0.176$, $\epsilon_2=12.4$ and $\epsilon_1=11.71$,
(which according to \eqref{e1_x} corresponds to $x=0.3$).
As can be seen from Fig. \ref{fig:Eb_B_d}(a)
$E^B_{D^0}$ is less than zero
only for small $d$
and sufficiently high magnetic field.
In this range of parameters
the complex $(D^+,X)$ is unstable 
due to dissociation process \eqref{disD}.
In turn,
Fig. \ref{fig:Eb_B_d}(b) shows that
it is unstable due to dissociation process \eqref{disX}
if $d$ is sufficiently large.
This critical distance decreases with increasing field
--- it may be due to the fact that in a magnetic field
the Coulomb interaction energy in $X$ is growing relatively quickly
while the ionized donor (because of repulsive potential for hole)
prevents such rapid growth of this energy in bound exciton.

Summarizing,
in the magnetic field 
for $x=0.3$ the complex is stable 
only for respectively small $d$ (the smaller the greater is the field)
excluding the range (small $d$ and big $B$)
in which
complex can dissociate into $D^0$ and~$h$.

\section{Conclusions}

In the present work we have calculated, using Hartree approximation,
the energy of exciton bound by distant ionized donor in 2D QW
for dielectric constant of QW material equal and greater 
than the dielectric constant of a barrier, where the donor was located.
In the latter case it turned out that in order to bound exciton 
the donor has to be shifted from the QW on certain distance.
Dependence of this distance on the dielectric constants has been calculated.
Moreover, we have also studied stability of created complex
depending on the value of dielectric constant of the barrier material 
and on the magnetic field.
%
%

Despite the fact that the potential of shifted donor
resembles (in the QW plane) the potential of type II quantum dot
it should be however possible to distinguish between the two
due to the mentioned lack of stability
of donor bound exciton in the growing magnetic field.




\end{document}